\documentclass[10pt]{iopart}

\usepackage{graphicx}
\expandafter\let\csname equation*\endcsname\relax
\expandafter\let\csname endequation*\endcsname\relax
\usepackage[fleqn]{amsmath}
\usepackage{hyperref}
\usepackage{cleveref}
\usepackage{amsfonts}
\usepackage{amssymb}
\usepackage{mathrsfs}
\usepackage[utf8]{inputenc}
\usepackage[T1]{fontenc}
\crefname{equation}{}{}
\crefname{figure}{figure}{}
\usepackage[numbers]{natbib}
\usepackage{xspace}

\newcommand{\gke}{\textsc{Gkeyll}\xspace}
\newcommand{\dx}[1]{\textnormal{d}#1}
\newcommand{\pderiv}[2]{
\frac{\partial #1}{\partial #2}
}
\newcommand{\deriv}[2]{
\frac{\dx{#1}}{\dx{#2}}
}
\newcommand{\pderivInline}[2]{
\partial #1/\partial #2
}
\renewcommand{\vec}[1]{\ensuremath{\mbox{\boldmath$ {#1} $}}} 
\newcommand{\uv}[1]{\ensuremath{\mathbf{\hat{#1}}}} 
\newcommand{\uvg}[1]{\boldsymbol{\hat{\boldsymbol{#1}}}}

\let\ccdot=\cdot
\renewcommand{\cdot}{\boldsymbol{\ccdot}}

\begin{document}

\title[Reduction of transport due to magnetic shear in the scrape-off layer]{Reduction of transport due to magnetic shear in gyrokinetic simulations of the scrape-off layer}

\author{N. R. Mandell$^{1,2}$, G. W. Hammett$^3$, A. Hakim$^3$, M. Francisquez$^3$}
\address{$^1$ MIT Plasma Science and Fusion Center, Cambridge, MA, 02139, USA}
\address{$^2$ Department of Astrophysical Sciences, Princeton University, Princeton, NJ 08543, USA}
\address{$^3$ Princeton Plasma Physics Laboratory, Princeton, NJ 08543, USA}
\ead{nrm@mit.edu}

\date{\today}

\begin{abstract}
The effect of varying magnetic shear on scrape-off layer turbulence and profiles is studied via electromagnetic gyrokinetic simulations of a helical scrape-off layer model. We develop a model helical geometry with magnetic shear and a corresponding field-aligned coordinate system, which is used for simulations with the \gke code. We find that perpendicular transport is reduced in cases with stronger shear, resulting in higher peak particle and heat fluxes to the endplates. Electromagnetic effects slightly increase transport in strong shear cases.
\end{abstract}


\submitto{\PPCF}

\ioptwocol

\section{Introduction}

Developing a quantitative understanding of the mechanisms that control turbulence and transport in the tokamak edge and scrape-off layer (SOL) is important for the success of fusion. This region provides the boundary condition for the hot core and is also responsible for handling the power exhaust from the core. Mitigation strategies for the large heat loads in the SOL must be careful to avoid degrading the high core temperatures required to sustain a burning plasma \citep{carter2020}. Insights from theory and first-principles numerical modeling are important for solving these challenging issues. Significant progress has been made in modeling the edge and SOL using fluid \citep{xu2008boundary, ricci2012simulation,halpern2016a,tamain2010tokam,zhu2018,stegmeir2018} and gyrokinetic \citep{ku2009full,ku2016new,korpilo2016gyrokinetic,dorf2016continuum,shi2019,mandell2020,pan2018,michels2021} models.

The geometry of the SOL in a diverted plasma can be challenging for turbulence modeling because field-aligned coordinate systems \citep{beer1995field} (which are standard for core turbulence simulations) become singular due to the presence of the X point. While several gyrokinetic codes are now capable of handling X-point geometry by departing from field-aligned coordinates \citep{ku2009full,dorf2016continuum,michels2021}, implementation of X-point geometry in our efforts is left to future work. 
Instead, we aim here to study the fundamental impact of magnetic shear on scrape-off layer turbulence, since strong magnetic shear in the vicinity of the X point is a key feature of realistic SOL geometries. The local shear can approach $\hat{s}= (r/q)d q/dr= 40$ one gyroradius away from the X point \citep{myra2000}. This results in significant radial shearing of eddies as one moves along the field line. In field-aligned coordinates, this is manifested by an increase in the radial mode number, $k_x \propto \hat{s}$, which in the large-aspect-ratio circular limit is given by $k_x = k_y \hat{s} (\theta-\theta_0)$, where $\theta_0$ is the ballooning angle (the poloidal angle where the local $k_x = 0$). The effects of magnetic shear on blob dynamics have been studied by \citep{dippolito2011} and \citep{krasheninnikov2008}. They showed that strong shear can reduce the blob velocity by allowing cross-field currents to close through the thin sheared part of the blob, shorting out the blob polarization.  
This is a similar physical mechanism to the one that gives the well-known stabilizing effect of magnetic shear in ballooning mode theory for closed-field-line geometries \citep{connor1978}. Further, \citep{myra2000} showed that strong magnetic shear localized near X-points produces resistive X-point (RX) modes in both the (confined) edge and the SOL. These modes become disconnected across the X-point region, producing a fast-growing instability. As a result, the modes become disconnected from their boundary conditions, resulting in similar dynamics for both the edge and SOL. RX modes were found to be the dominant instability in L-mode conditions in \citep{xu2000a,xu2000b}. A theoretical investigation of strong magnetic shear near X points is also given in \cite{stoltzfus2009}. Note that in our studies we will not have localized strong magnetic shear regions because we do not have X points, so we do not expect to see RX modes.

The \gke code\footnote{\url{https://gkeyll.readthedocs.io}} has recently become the first code to demonstrate the capability to simulate electromagnetic gyrokinetic turbulence on open magnetic field lines. Electromagnetic results have been shown in a simple helical SOL geometry with no magnetic shear \citep{mandell2020,hakim2020a, mandell2021,mandell2022a, mandell2022b}. Previous electrostatic \gke results also used a helical geometry without shear, including studies of an NSTX-like helical SOL \citep{shi2019} and the Texas Helimak \citep{bernard2020}. In this work we extend the geometry in \gke to include magnetic shear and all other geometrical factors arising from using a non-orthogonal field-aligned coordinate system. 
We then include magnetic shear in full-$f$ electromagnetic gyrokinetic simulations with \gke for the first time, taking a helical SOL geometry with magnetic shear and parameters that roughly approximate the SOL of the National Spherical Torus Experiment (NSTX). 

The remainder of this paper is organized as follows: In Section 2, we describe the electromagnetic gyrokinetic model used for the simulations. We describe the helical SOL configuration, including magnetic shear, in section 3, along with details about the field-aligned coordinate system that we choose for this configuration. Section 4 presents the results of a parameter scan of the magnetic shear, showing electrostatic and electromagnetic simulations. The main result is that magnetic shear reduces transport, resulting in steeper profiles in cases with more shear. Conclusions are given in Section 5.

\section{Gyrokinetic model}
We model turbulence by solving the full-$f$ electromagnetic gyrokinetic (EMGK) system in the long-wavelength (drift-kinetic) limit. The gyrokinetic equation describes the evolution of the guiding-center distribution function $f_s=f_s(\vec{R},v_\parallel,\mu;t)$ for species $s$, where $\vec{R}=(x,y,z)$ is the guiding-center position, $v_\parallel$ is the parallel velocity, and $\mu = m_s v_\perp^2/(2B)$ is the magnetic moment. In conservative form we have
\begin{align}
&\pderiv{(\mathcal{J}f_s)}{t} + \nabla{\cdot}( \mathcal{J} \dot{\vec{R}} f_s) + \pderiv{}{v_\parallel}\left(\mathcal{J}\dot{v}^H_\parallel f_s\right) 
\notag \\&\quad
- \pderiv{}{v_\parallel}\left(\mathcal{J}\frac{q_s}{m_s}\pderiv{A_\parallel}{t} f_s \right) 
= \mathcal{J} C[f_s] + \mathcal{J} S_s, \label{emgk} 
\end{align}
where the nonlinear phase-space characteristics are given by
\begin{gather}
    \dot{\vec{R}} = \frac{\vec{B^*}}{B_\parallel^*}v_\parallel + \frac{\uv{b}}{q B_\parallel^*}\times\left(\mu\nabla B + q \nabla \Phi\right), \label{eomR0}\\
    \dot{v}_\parallel = -\frac{\vec{B^*}}{m B_\parallel^*}{\cdot}\left(\mu\nabla B + q \nabla \Phi\right)-\frac{q}{m}\pderiv{A_\parallel}{t}, \label{eomV0}
\end{gather}
with $\Phi$ the electrostatic potential, $A_\parallel$ the parallel magnetic vector potential. Collisions $C[f_s]$ and sources $S_s$ are included on the right-hand side of \cref{emgk}. Here, $B_\parallel^*=\uv{b}\vec{\cdot} \vec{B^*}$ is the parallel component of the effective magnetic field $\vec{B^*}=\vec{B}+(m_s v_\parallel/q_s)\nabla\times\uv{b} + \delta \vec{B}$, where $\vec{B} = B \uv{b}$ is the equilibrium magnetic field and $\delta \vec{B} = \nabla\times(A_\parallel \uv{b})$ is the perturbed magnetic field, neglecting compressional magnetic fluctuations. The Jacobian of the gyrocenter coordinates is $\mathcal{J}=B_\parallel^*/m_s$, and we make the approximation $\uv{b}\cdot\nabla\times\uv{b}\approx0$ so that $B_\parallel^*\approx B$. The species charge and mass are $q_s$ and $m_s$, respectively. Note that we use the ``symplectic'' formulation \citep{brizard2007foundations} of electromagnetic gyrokinetics here, so that $v_\parallel$ is used as a coordinate and the inductive component of the parallel electric field, $\pderivInline{A_\parallel}{t}$, appears explicitly in \cref{emgk}.

The electrostatic potential is determined by the quasi-neutrality condition in the long-wavelength limit, which takes the form of the Poisson equation:
\begin{equation}
    -\nabla \cdot \left(\epsilon_\perp \nabla_\perp \Phi\right) = \sum_s q_s \int  f_s\, \dx{^3\vec{v}}, \label{poisson1}
\end{equation}
with $\dx{^3 \vec{v}}=2\pi \dx{} v_\parallel \dx{} \mu \mathcal{J}$ and
\begin{equation}
    \epsilon_\perp = \sum_s \frac{m_s n_{0s}}{B^2}.
\end{equation}
Here, we use a linearized polarization density $n_0$ that we take to be a constant in time, which is consistent with neglecting a second-order $E\times B$ energy term in the Hamiltonian. While the validity of this  Boussinesq-type approximation in the SOL can be questioned due to large density fluctuations (and we plan to eventually improve on this approximation), a linearized polarization density is commonly used for computational efficiency \citep{ku2018fast,shi2019}. The magnetic vector potential is determined by the parallel Amp\`ere equation,
\begin{equation}
    -\nabla_\perp^2 A_\parallel = \mu_0 \sum_s q_s \int  v_\parallel f_s\,\dx{^3\vec{v}}. \label{ampere1}
\end{equation}
Note that one can take the time derivative of \cref{ampere1} to obtain an exact Ohm's law that can be used to solve for $\pderivInline{A_\parallel}{t}$ directly. For more details on the electromagnetic gyrokinetic system and our implementation of the system in the \gke code, see \citep{mandell2020,mandell2021}.

To model the effect of collisions we use a conservative Lenard--Bernstein (or Dougherty) collision operator
\citep{Lenard1958,Dougherty1964},
\begin{align} \label{eq:GkLBOEq}
\mathcal{J}C[f_s] &= \sum_r\nu_{sr}\left\lbrace\pderiv{}{v_\parallel}\left[\left(v_\parallel - u_{\parallel sr}\right)\mathcal{J} f_s+v_{tsr}^2\pderiv{(\mathcal{J} f_s)}{v_\parallel}\right] 
\right. \notag\\ &\qquad \left.
+\pderiv{}{\mu}\left[2\mu \mathcal{J} f_s+2\mu\frac{m_s}{B}v_{tsr}^2\pderiv{(\mathcal{J} f_s)}{\mu}\right]\right\rbrace,
\end{align}
where like-species collisions use $u_{\parallel sr}=u_{\parallel s}$, $v_{tsr}=v_{ts}$ and these quantities are given by
\begin{align}
    &n_s u_{\parallel s} = \int v_\parallel f_s \,\dx{^3\vec{v}}, \\
    &n_s u_{\parallel s}^2 + 3 n v_{ts}^2 = \int  \left(v_\parallel^2 + 2\mu B/m_s\right)f_s\,\dx{^3\vec{v}},
\end{align}
with $n_s = \int f_s\,\dx{^3\vec{v}}$ (see \citep{francisquez2020} for more details). Cross-species collisions among electrons and ions are also modeled \citep{francisquez2021}.
This collision operator contains the effects of drag and pitch-angle scattering, and it conserves number, momentum and energy density. Consistent with our present long-wavelength treatment of the gyrokinetic system, finite-Larmor-radius effects are ignored. Note that in this model collision operator, the collision frequency $\nu$ is velocity-independent, i.e. $\nu\neq\nu(v)$.

\section{Helical SOL configuration with magnetic shear} \label{sec:heli-geo}
\begin{figure*}
    \centering
    \includegraphics[width=.6\textwidth]{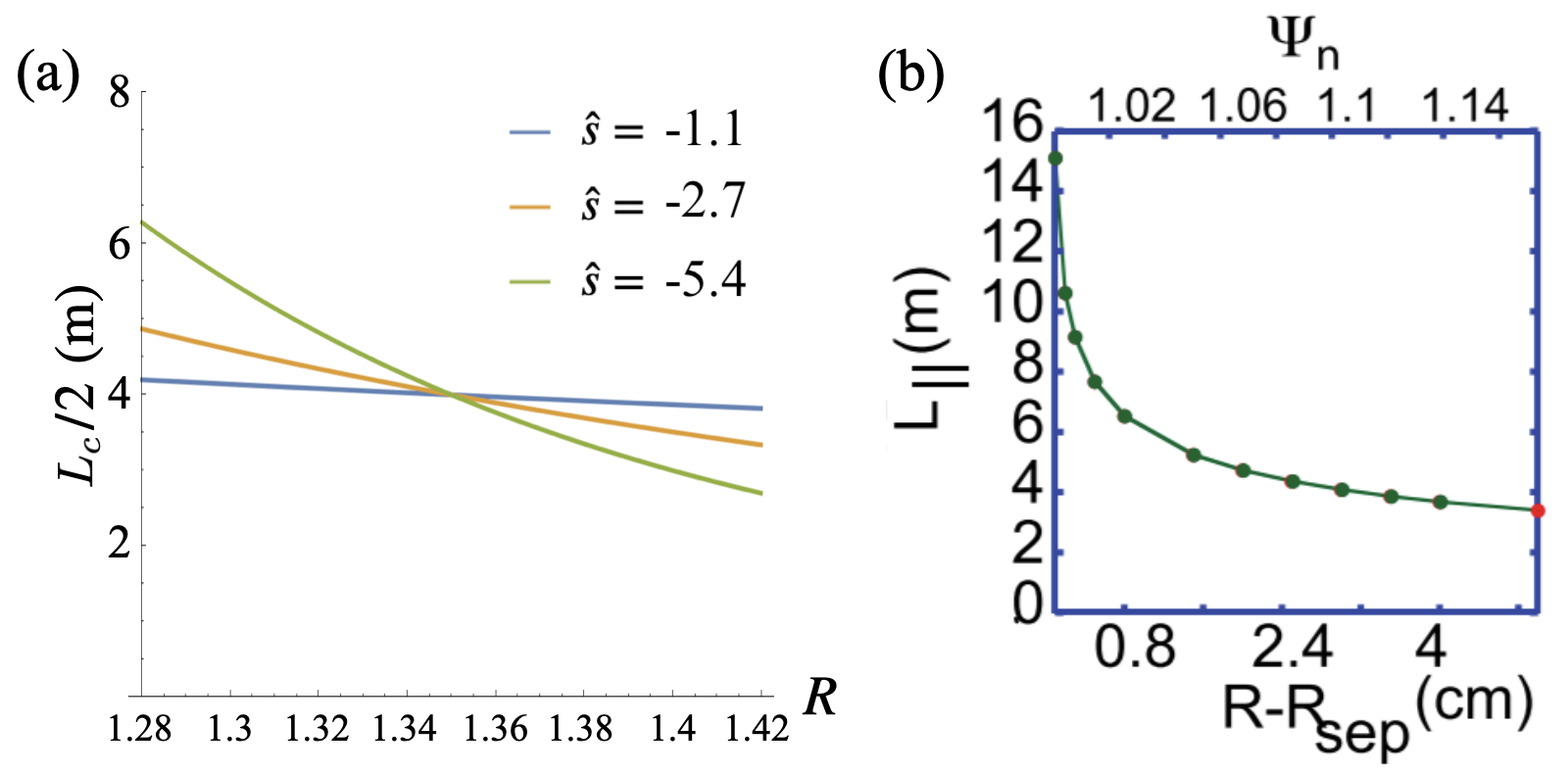}
    \caption[Connection length, $L_c=H B/B_v$, as a function of radius for several values of magnetic shear, $\hat{s}$, using NSTX-like parameters.]{$(a)$ Connection length, $L_c/2=H B/B_v/2$, as a function of radius for several values of magnetic shear, $\hat{s}$, using NSTX-like parameters: $R_0=0.85$ m, $x_0=R_0+a=1.35$ m, $B_0=0.5$ T, $H=2.4$ m. Here we choose $B_{v0}$ so that $L_c(x_0)=8$ m, which results in $B_{v0}\approx 0.1$ T. $(b)$ Connection length from the NSTX experiment, $L_\parallel$, defined as parallel length from midplane to lower divertor plate. Figure $(b)$ reprinted from Boedo \emph{et. al.}, \emph{Phys. Plasmas} {\bf 21} 42309 (2014) \cite{boedo2014}, with the permission of AIP Publishing.}
    \label{fig:nstx-heli-Lc}
\end{figure*}
\begin{figure*}
    \centering
    \includegraphics[width=.5\textwidth]{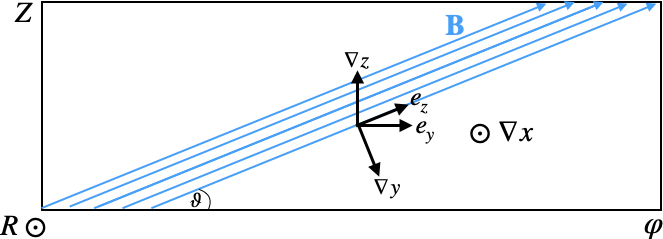}
    \caption[Diagram of field-aligned coordinate basis vectors in the $(\varphi,Z)$ plane for a helical magnetic geometry.]{Diagram of field-aligned coordinate basis vectors in the $(\varphi,Z)$ plane for a helical magnetic geometry. Note that vector magnitudes are not drawn to scale.}
    \label{fig:heli-geo}
\end{figure*}

The helical field in a simple magnetized torus (SMT) is given in cylindrical  $(R,\varphi,Z)$ coordinates (where $\varphi$ is \textit{counter-clockwise} viewed from above) by
\begin{equation}
    \vec{B} = B_\varphi \uvg{\varphi} + B_v \uv{Z} = \frac{B_0 R_0}{R} \uvg{\varphi} + B_v \uv{Z}.
\end{equation}
Note we can also write this as
\begin{equation}
    \vec{B} = \nabla\Psi\times\nabla\varphi + B_0 R_0 \nabla \varphi,
\end{equation}
with $\Psi = R^2 B_v/2$ the vertical magnetic flux function (analogous to the poloidal flux in a tokamak). The field line pitch varies with radius, and can be expressed via
\begin{equation}
    q(R) = \frac{H B_\varphi}{\pi R B_v} = \frac{B_0 R_0 H}{\pi R^2 B_v},
\end{equation}
where $q$ is analogous to the safety factor in a tokamak, and $H$ is the vertical height between the bottom and top end-plates where the field lines terminate. Here we have identified the poloidal length of the bad curvature region of a tokamak, $\pi r$ (with $r$ the minor radius) with the height $H$, so that our expression for $q$ corresponds to the standard cylindrical tokamak expression $q = (r/R)(B_\phi/B_\theta)$ with $B_\phi$ and $B_\theta$ the toroidal and poloidal components of the magnetic field in a tokamak.
Similarly, we can define the magnetic shear as
\begin{equation}
    \hat{s} = \frac{H}{\pi q}\frac{B_\varphi}{B}\deriv{q}{R},
\end{equation}
which is analogous with the standard definition in a tokamak, $\hat{s}=(r/q)\dx{q}/\dx{r}$. Here, note that our definition of $q$ is different by a factor of 2 from the expressions of \citep{perez2006}, who effectively identify $H$ with $2\pi r$ in a tokamak. We also used different definitions of $q$ and $\hat{s}$ in \citep{mandell2020}; we changed the definitions here for better correspondence with standard tokamak geometry.

If we allow the vertical field to have some simple radial dependence via $B_v=B_v(R) = B_{v0}(R/x_0)^n$, with $x_0$ some major radius of interest, the shear is
\begin{equation}
    \hat{s} = -\frac{H}{\pi R}\frac{B_\varphi}{B}(n+2).
\end{equation}
When $B_v=$ const ($n=0$, as is the case for standard SMTs like the Helimak), we have $\hat{s}=-2HB_\varphi/(\pi B R)$. The connection length is given by $L_c = H B/B_v$, which in general varies with radius. In \cref{fig:nstx-heli-Lc}$(a)$ we plot the connection length $L_c$ as a function of radius for several values of $\hat{s} = \{-1.1,-2.7,-5.4\}$ (corresponding to $n=\{0,3,8\}$) with NSTX-like geometrical parameters (and evaluating $\hat{s}$ at $R=x_0$). 

Note that here all quantities, including the field line pitch and the magnetic shear, are still constant \emph{along} the field lines. This is in contrast to a real tokamak SOL, where the pitch and shear can vary significantly along the field lines, especially near the X-point due to flux expansion. The field line pitch at the midplane is also nearly constant with radius in a real tokamak SOL, while here we have the pitch varying with radius. Nonetheless, this simple geometry will still allow us to study some of the effects of magnetic shear. Comparing to the actual connection length in the NSTX experiment shown in \cref{fig:nstx-heli-Lc}$(b)$ (with the figure adapted from Fig. 2 in \citep{boedo2014}), we see that the variation in $L_c$ with radius as $|\hat{s}|$ increases is approaching the variation of the connection length in the experiment. However, in the experiment the connection length varies even more than in the $\hat{s}=-5.4$ case over a shorter radial length, changing by a factor of almost four over a few centimeters.

\subsection{Field-aligned coordinate system in helical geometry}
In order to take advantage of the fact that turbulent structures are much more elongated along the field line than perpendicular to it ($k_\parallel \ll k_\perp$), we adopt a field-aligned coordinate system, which we will denote by $(x,y,z)$. To do this, we write the background magnetic field in Clebsch-like form as
\begin{equation}
    \vec{B} = \mathcal{C}(x) \nabla x \times \nabla y,
\end{equation}
where $x$ and $y$ are coordinates perpendicular to the background field, with $x$ a radial-like coordinate, and $y$ a field-line-labeling coordinate. 
This can be achieved by choosing the coordinates to be\footnote{Alternatively, we could have chosen the $z$ coordinate to measure distance along the field line, with $z = Z/\sin\vartheta = Z B/B_v$. In this approach, the $z$ domain extent is given by the connection length; however, the connection length can vary with radius, so the $z$ domain extent would also need to vary with radius. Choosing the vertical height $Z$ for the $z$ coordinate does not have this issue (assuming the vertical height between the top and bottom end plates does not change with radius), so this is the approach that we take in this work.}
\begin{align}
    x &= R, \\
    z &= Z, \\
    y &= x_0\left(\varphi - \frac{\pi qZ}{H}\right)=x_0\left(\varphi - \frac{B_\varphi Z}{B_v R} \right). \label{heli-coord}
\end{align}
The resulting gradient basis vectors are
\begin{align}
    \nabla x &= \uv{R}, \\
    \nabla y &= -\frac{\pi x_0 z}{H}\deriv{q}{x}\uv{R} + \frac{x_0}{x}\uvg{\varphi} - \frac{\pi x_0 q}{H} \uv{Z} 
    , \\
    \nabla z &= \uv{Z}, \label{gradvec}
\end{align}
so that
\begin{equation}
    \nabla x \times \nabla y  = \frac{x_0}{x} \left(\frac{\pi qx}{H}\uvg{\varphi} + \uv{Z}\right) =  \frac{x_0}{x} \left(\frac{B_\varphi}{B_v} \uvg{\varphi} + \uv{Z}\right). 
\end{equation}
Now taking $\mathcal{C}(x) = B_v x/x_0$, we obtain the correct form of the field,
\begin{equation}
    \vec{B} = \frac{B_v x}{x_0} \nabla x \times \nabla y = B_\varphi\uvg{\varphi} + B_v\uv{Z}.
\end{equation}
A diagram of the field-aligned basis vectors in the $(\varphi,Z)$ plane is shown in \cref{fig:heli-geo}. 

The Jacobian of the field-aligned coordinate system is
\begin{equation}
    J = [(\nabla x\times \nabla y) \cdot \nabla z]^{-1} = \frac{x}{x_0}.
\end{equation}
Note that we can write
\begin{equation}
    \nabla y = \frac{B}{JB_v}\left(\hat{s}\theta \uv{R} + \frac{B_v}{B}\uvg{\varphi} - \frac{B_\varphi}{B} \uv{Z}\right),
\end{equation}
where we have identified $\theta = z \pi/H$, so that
\begin{equation}
    g^{yy} = |\nabla y|^2 = \left(\frac{B}{J B_v}\right
    )^2(1 + \hat{s}^2 \theta^2).
\end{equation}
This corresponds to the standard definition of $g^{yy}=1+\hat{s}^2\theta^2$ in a circular large-aspect-ratio tokamak, which confirms that we have made the appropriate definition of $\hat{s}$.

For more details about our implementation of the gyrokinetic system in a field-aligned coordinate system, see \citep{mandell2021}. 

\begin{figure*}
    \centering
    \includegraphics[width=.8\textwidth]{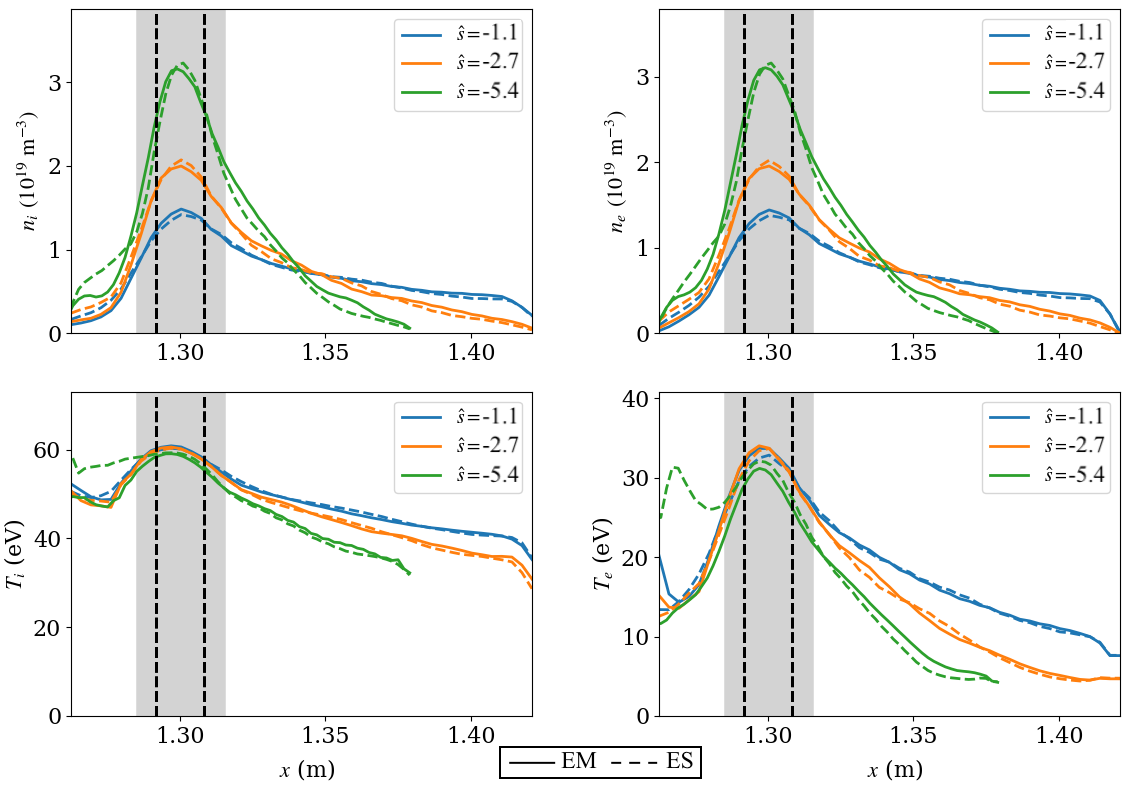}
    \caption[Midplane density and  temperature profiles for several values of magnetic shear, $\hat{s}$.]{Time- and $y-$averaged midplane density and  temperature profiles for several values of magnetic shear, $\hat{s}$, with ion quantities on the left and electron quantities on the right. Both electromagnetic (EM, solid lines) and electrostatic (ES, dashed) results are shown, with electromagnetic effects not having a significant effect on the profiles. The profiles steepen as $|\hat{s}|$ increases and the geometry becomes more sheared.}
    \label{fig:shat-profiles}
\end{figure*}

\section{Results}
In this section we present gyrokinetic simulations in the helical configuration with magnetic shear described above. The simulations were performed with the EMGK module of the \gke plasma simulation framework. This module employs a discontinuous Galerkin (DG) discretization scheme for the EMGK system that conserves energy (in the continuous-time limit) and avoids the Amp\`ere cancellation problem \citep{mandell2020}. 

We perform a scan of the magnetic shear parameter, taking $\hat{s}=\{-1.1,-2.7,-5.4\}$, with the geometry becoming more sheared as $|\hat{s}|$ increases. In each case, we perform both an electrostatic simulation and a simulation including electromagnetic effects. We use NSTX-like geometry parameters: $R_0=0.85$ m, $x_0=R_0+a=1.35$ m, $B_0=0.5$ T, $H=2.4$ m, and we choose $B_{v0}=B_v(x_0)$ so that $L_c(x_0)=8$ m, resulting in $B_{v0}\approx 0.1$ T. The resulting connection length as a function of radius is shown in \cref{fig:nstx-heli-Lc}. The simulation domain extents are taken to be $1.26 \leq x \leq 1.42$ m ($L_x \approx 56\rho_{\mathrm{s}0}$) for the $\hat{s}=-1.1$ and $\hat{s}=-2.7$ case, and $1.26 \leq x \leq 1.38$ ($L_x\approx42\rho_{s0}$) for the $\hat{s}=-5.4$, where we can use a smaller radial domain due to lower transport levels. The domain extents in the binormal and parallel directions, respectively, are $-0.485 \leq y \leq 0.485$ m ($L_y \approx 100\rho_{\mathrm{s}0} H/L_c(x_0)$) and $-H/2 \leq z \leq H/2$ in all cases. In velocity space, we use $-4v_{ts}\leq v_\parallel \leq 4 v_{ts}$ and $0\leq\mu\leq6T_{0}/B_0$, where $v_{ts}=\sqrt{T_{0}/m_s}$ with $T_0=40$ eV and $B_0=B_\text{axis}R_0/R_c$.

The simulations use piecewise-linear ($p=1$) basis functions, with $(N_x,N_y,N_z,N_{v_\parallel},N_\mu)=(48,96,16,10,5)$ the number of cells in each dimension for the $\hat{s}=-1.1,-2.7$ cases and $(30,96,16,10,5)$ cells for the $\hat{s}=-5.4$ case (which still gives slightly finer radial resolution than the other cases because the radial domain extent has also been reduced for $\hat{s}=-5.4$). Note that for $p=1$ DG, one should double each of these numbers to obtain the equivalent number of grid-points for comparison with standard grid-based gyrokinetic codes. All other parameters are the same as used in the base case ($\hat{n}=1$) from \citep{mandell2022a}, including the source power $P_\mathrm{src} = P_\mathrm{SOL} L_y/(2\pi R_c)= 0.62$ MW and the source profile with a Gaussian peak at $x=1.3$ m. Boundary conditions are also the same as in \citep{mandell2022a}, with Dirichlet boundary conditions $\Phi=A_\parallel=0$ on the radial walls, periodic boundary conditions in the $y$ direction, and conducting-sheath boundary conditions in the $z$ direction. Our present sheath boundary condition does not account for the grazing incidence angle of the field lines on the endplates, which modifies the sheath physics \citep{chodura1982} and the dynamics \citep{geraldini2017}. Instead we simply assume perpendicular incidence angle (as if the endplates were tilted appropriately at every incidence point) and leave more realistic Chodura sheath boundary conditions to future work.

\begin{figure*}
    \centering
    \includegraphics[height=9cm]{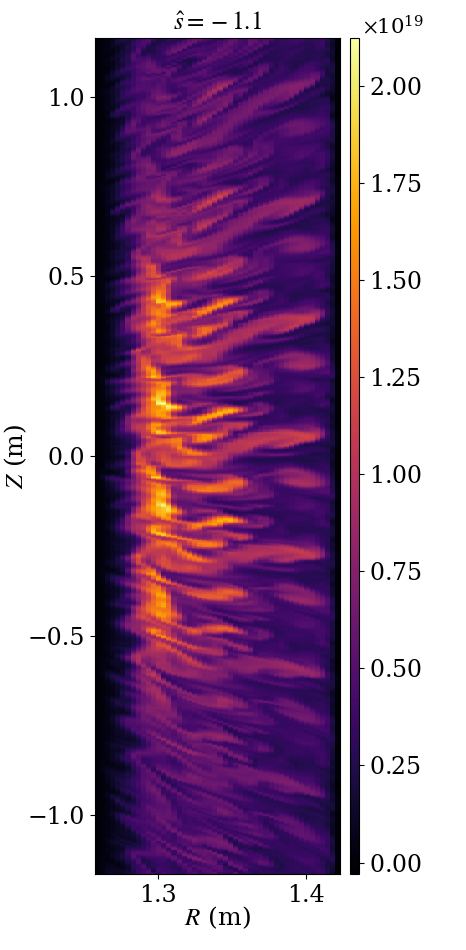}
    \includegraphics[height=9cm]{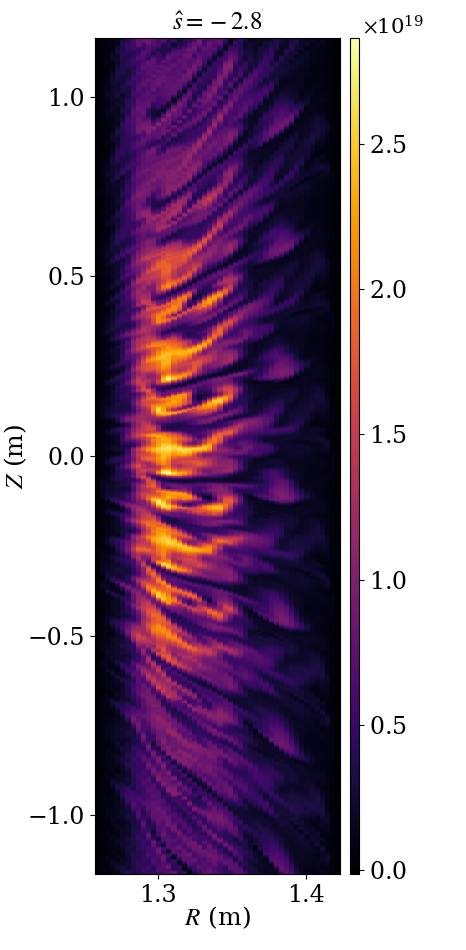}
    \includegraphics[height=9cm]{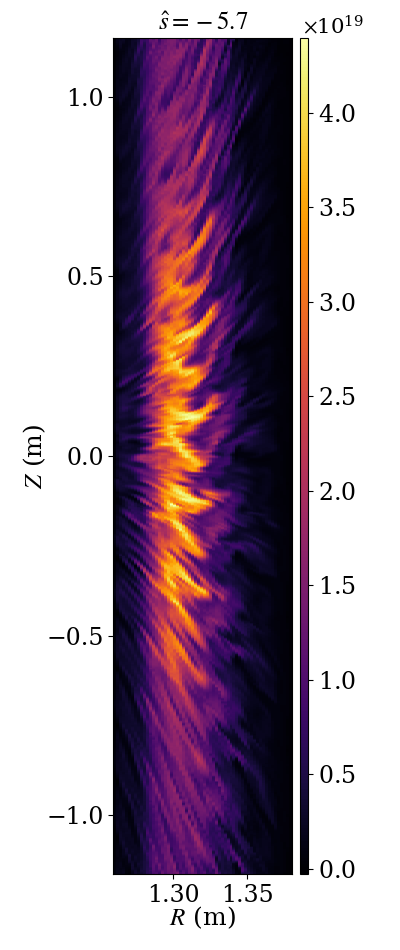}
    \caption{Snapshots of electron density (in m$^{-3}$) mapped to the $(R,Z)$ plane at $t=800\ \mu$s for the $\hat{s}=-1.1,-2.7,-5.4$ electromagnetic cases.}
    \label{fig:shat-snapshots}
\end{figure*}

In \cref{fig:shat-profiles} we show time- and $y-$averaged density and temperature radial profiles for each case. Electromagnetic cases are shown with solid lines, while the electrostatic cases are shown dashed. The profiles for the $\hat{s}=-1.1$ case are similar to the profiles from the base case in \citep{mandell2022a}, which used a simplified geometry neglecting magnetic shear and assumed a constant connection length. This is somewhat expected since \cref{fig:nstx-heli-Lc} shows that the connection length in the $\hat{s}=-1.1$ case varies little over the domain. As we move to more sheared geometries, the density profiles steepen, with the peak midplane density more than doubling between the $\hat{s}=-1.1$ and $\hat{s}=-5.4$ cases. In all cases there is not much effect on the midplane profiles from including electromagnetic perturbations. 

\begin{figure*}
    \centering
    \includegraphics[width=.8\textwidth]{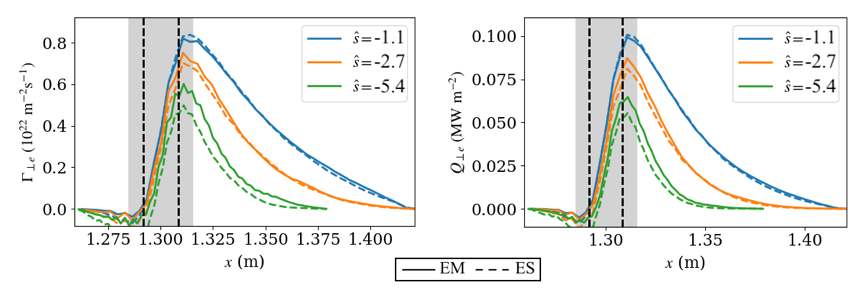}
    \caption[Radial electron $E\times B$ particle flux near the midplane for several values of $\hat{s}$.]{Radial electron $E\times B$ particle flux (left) and heat flux (right) near the midplane, for both electromagnetic (solid) and electrostatic (dashed) cases. The cross-field transport decreases as the geometry becomes more sheared at larger $|\hat{s}|$, consistent with the steeper profiles seen in \cref{fig:shat-profiles}. Electromagnetic effects result in slightly increased transport in the stronger shear cases.}
    \label{fig:shat-exb-flux}
\end{figure*}

Snapshots of the electron density mapped to the $(R,Z)$ plane for each of the electromagnetic cases are shown in \cref{fig:shat-snapshots}. We can clearly see that at large $|Z|$ the turbulent structures are being tilted due to magnetic shear, with more tilting as $|\hat{s}|$ increases.
Near the midplane at $Z=0$, the $\hat{s}=-1.1$ case looks similar to the cases from \citep{mandell2022a}, with blobs moving radially outwards and structures extending most of the width of the radial domain. On the other hand, the $\hat{s}=-5.4$ case clearly shows evidence of reduced radial transport.

\begin{figure*}
    \centering
    \includegraphics[width=.9\columnwidth]{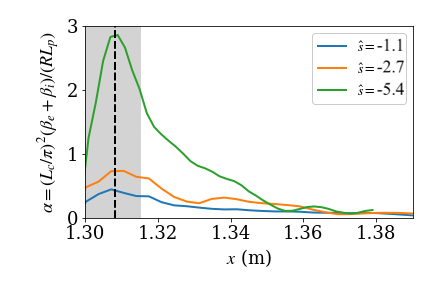}
    \caption{Radial profiles of the ballooning stability parameter $\alpha = \gamma_\mathrm{int}^2/(k_{\parallel 0}^2 v_A^2) = (L_c/\pi)^2(\beta_e+\beta_i)/(R L_p)$ at the midplane for each of the electromagnetic cases.}
    \label{fig:alpha}
\end{figure*}

\begin{figure*}
    \centering
    \includegraphics[width=.8\textwidth]{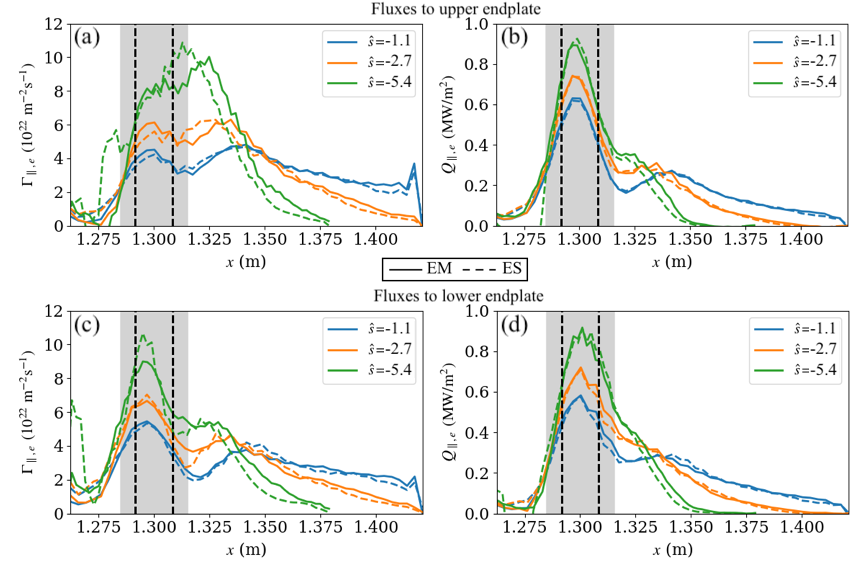}
    \caption[Electron particle and heat fluxes to the end plates for several values of $\hat{s}$.]{Time-averaged electron particle (left) and heat (right) fluxes to the upper (top row) and lower (bottom row) end plates. The peak fluxes increase as $|\hat{s}|$ increases, consistent with less cross-field transport upstream.}
    \label{fig:shat-end-flux}
\end{figure*}

We can quantify this by computing the radial electron $E\times B$ particle flux as
\begin{equation}
    \Gamma_{\perp e} = \langle \tilde{n}_e \tilde{v}_r \rangle +  \langle \tilde{u}_{\parallel e} \tilde{b}_r \rangle, \label{gamperp}
\end{equation}
and the radial electron $E\times B$ heat flux as
\begin{equation}
    Q_{\perp e} = \langle \tilde{p}_e \tilde{v}_r \rangle + \langle \tilde{q}_{\parallel e} \tilde{b}_r \rangle. \label{qperp}
\end{equation}
Here, the first term is the contribution from the $E\times B$ drift, with $v_r = E_r/B = -(1/B)\pderivInline{\Phi}{y}$, $n_e$ the electron density, $p_e$ the electron pressure. The second term is the flux due to magnetic flutter (for the electromagnetic cases only), with $b_r = (1/B)\pderivInline{A_\parallel}{y}$, $u_{\parallel e}$ the electron parallel flow, and $q_{\parallel e}$ the electron parallel heat flux. The tilde indicates the fluctuation of a time-varying quantity, defined as $\tilde{F}=F - \bar{F}$ with $\bar{F}$ the time average of $F$. The brackets $\langle F \rangle$ denote an average in $y$ and time.
We plot these quantities, evaluated near the midplane, in \cref{fig:shat-exb-flux} and confirm that indeed both particle and heat transport are reduced as $|\hat{s}|$ increases.

Interestingly, despite the fact that we observed little change in the midplane profiles due to electromagnetic effects in \cref{fig:shat-profiles}, we see here that for the cases with more shear, electromagnetic effects are resulting in slightly more radial transport. Electromagnetic effects were also observed to increase transport in the no-shear simulations of \citep{mandell2022a}, although in those cases, departure from the electrostatic limit required artificially increasing the source power to access a higher $\beta$ regime. Here we have kept the nominal experimental $P_{SOL}$ and yet we are still seeing the influence of electromagnetic effects on the radial transport. The key is that stronger magnetic shear results in steeper pressure profiles, which can strengthen interchange-ballooning instabilities via the ballooning stability parameter $\alpha = \gamma_\mathrm{int}^2/(k_{\parallel 0}^2 v_A^2)$, where $k_{\parallel 0}=\pi/L_c$ is the longest parallel half-wavelength on the domain. Due to sheath and finite radial-mode-width effects, however, there is not a sharp limit on $\alpha$ in the scrape-off layer \citep{mandell2022b}. Extension of the sheath-modified interchange-ballooning analysis in helical geometry presented in \citep{mandell2022b} to include magnetic shear is left to future work, but we do plot $\alpha = \gamma_\mathrm{int}^2/(k_{\parallel 0}^2 v_A^2) = (L_c/\pi)^2(\beta_e+\beta_i)/(R L_p)$ evaluated at the midplane in \cref{fig:alpha}, where here $k_{\parallel 0} = \pi / L_c$ itself varies with radius because of the variation of $L_c$. Taking the peak $\alpha$ value from each the case, we have a range $\alpha \approx 0.3 - 3$, which is in the range of measurements of $\alpha$ in the near SOL of present experiments \citep{labombard2008,eich2018,eich2021}.

\begin{figure*}
    \centering
    \includegraphics[width=.55\textwidth]{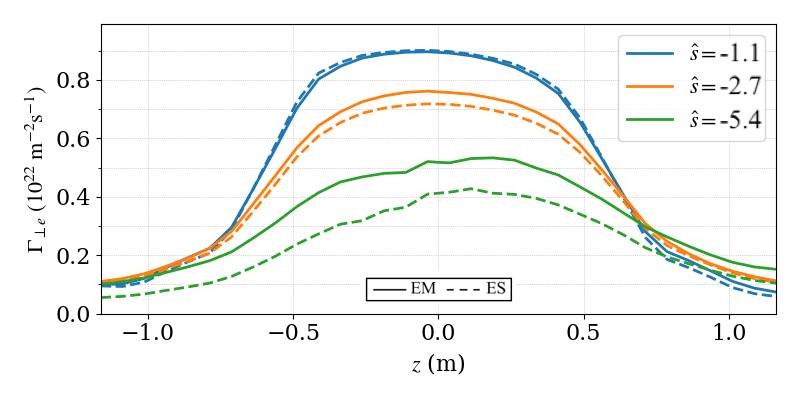}
    \caption[$\hat{s}$ scan: Electron radial $E\times B$ particle flux profiles along the field line.]{Time-averaged electron radial $E\times B$ particle flux profiles along the field line, taken just outside the source region at $x\sim 1.32$ m. In the $\hat{s}=-5.4$ cases there is a noticeable asymmetry in the profiles, with slightly more radial transport at $z>0$ than $z<0$. This is consistent with the differences in the particle fluxes to the bottom and top end plates shown in \cref{fig:shat-end-flux} $(a)$ and $(c)$.}
    \label{fig:shat-exb-flux-z}
    \vspace{.5cm}
    \centering
    \includegraphics[width=.8\textwidth]{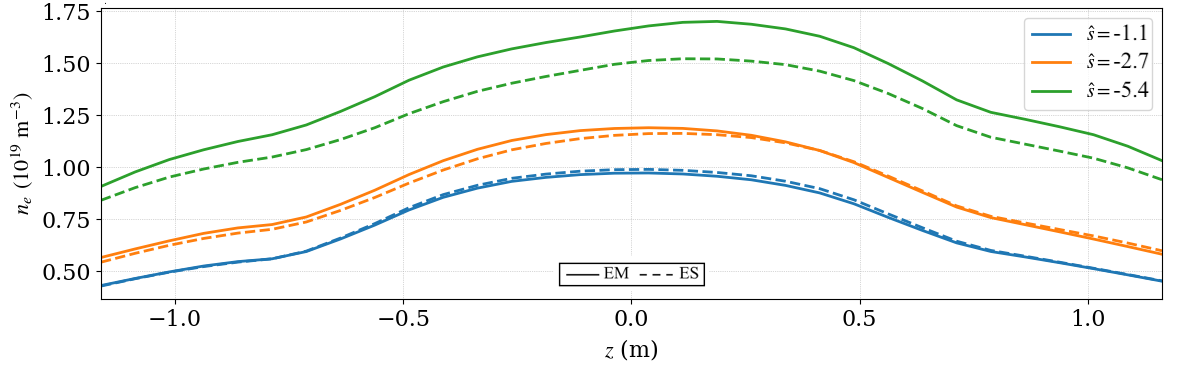}
    \caption[$\hat{s}$ scan: Electron density profiles along the field line.]{Time- and $y$-averaged electron density profiles along the field line, taken just outside the source region at $x\sim 1.32$ m. There is noticeable  asymmetric shift in the profiles in the $\hat{s}=-5.4$ cases, with slightly higher density near the top of the domain ($z=H/2=1.4$ m).}
    \label{fig:shat-ne-z}
\end{figure*}

As a result of weaker cross-field transport, the peak particle and heat fluxes to the end plates increase in the cases with more shear, as shown in \cref{fig:shat-end-flux}. Note that we have separately shown the fluxes to the upper (top row) and lower (bottom row) end plates. Despite the cross-field transport differences in the $\hat{s}=-5.4$ case observed above, there is not much broadening of the flux profiles due to electromagnetic effects, although the peak particle flux to the lower endplate is slightly reduced. There are also some interesting asymmetries in the flux profiles between the ends. This is especially apparent in the particle flux in the stronger shear cases, where there is a noticeable shift in the profiles to higher $x$ between the lower and upper plates. When we examine how the radial particle flux (taken just outside the source region near $x=1.32$ m) varies along the field line in \cref{fig:shat-exb-flux-z}, we see asymmetry as well. There is noticeably more radial transport at $z>0$ than $z<0$ for the $\hat{s}=-5.4$ cases. This is consistent with the radial shift in the endplate particle fluxes to higher $x$ between the lower and upper plates.

The average electron density also shows asymmetry along the field line, as shown in \cref{fig:shat-ne-z}. Here we have again evaluated the profiles just outside the source region near $x=1.32$ m. There is a slight shift in the profiles to higher $z$ in the $\hat{s}=-5.4$ cases. One possible reason for the asymmetry is the presence of a vertical component in the $E\times B$ and magnetic drifts.  For the $E\times B$ drift, this is 
\begin{align}
    \vec{v}_E\cdot\nabla z &= \frac{1}{J B}\left(b_x \pderiv{\Phi}{y} - b_y \pderiv{\Phi}{x} \right) \notag \\
    &= \frac{B_\varphi}{B^2}\left(-\pderiv{\Phi}{x} + \frac{x_0}{x}\frac{\pi}{H}\frac{B}{B_v}\hat{s} z \pderiv{\Phi}{y}\right)
\end{align}
These vertical drift terms were not present in the simulations in \citep{mandell2022a}, which used a simplified geometry in the $B_v\ll B$ limit. It has been suggested that vertical drifts are responsible for asymmetry between top and bottom profiles in the Helimak \citep{bernard2020}.

\section{Conclusions}

In this paper we have explored the effects of magnetic shear on scrape-off layer transport. We first introduced shear into a  helical SOL model geometry by taking the vertical field to be $B_v \sim R^n$ so that the magnetic shear is proportional to $-2 -n$. We next derived a field-aligned coordinate system in this geometry that was implemented into the \gke gyrokinetic code. We then performed electromagnetic and electrostatic gyrokinetic simulations in this geometry for several values of $\hat{s}$. The results showed that magnetic shear reduces perpendicular transport, resulting in steeper radial profiles and higher peak fluxes to the endplates in cases with more shear. We also found that electromagnetic effects enhanced transport slightly (compared to corresponding electrostatic simulations) in the strong shear cases. Asymmetry in the endplate flux profiles between the top and bottom plates was found to be consistent with asymmetry in perpendicular transport about the device midplane; this is believed to be due to vertical components of the $E\times B$ and magnetic drifts. 

Extension of this work to modeling of a shaped scrape-off layer in a limiter configuration is already underway \citep{mandell2021}. Of course, the eventual goal is to model diverted geometries with X points. This may require departure from field-aligned coordinate systems, at least in the immediate vicinity of the separatrix \citep{ku2009full,dorf2016continuum,michels2021}. Nonetheless, this work was an important step towards modeling realistic SOL geometry with \gke, which will enable closer comparison with experimental measurements.

\section*{Acknowledgements}
We would like to thank Tess Bernard, Petr Cagas, James Juno and other members of the \gke team for helpful discussions and support, including the development of the \texttt{postgkyl} post-processing tool which facilitated the creation of many figures in this paper.
Research support came from the U.S. Department of Energy: N.R.M. was supported by the DOE Fusion Energy Sciences Postdoctoral Research Program
administered by the Oak Ridge Institute for Science and Education (ORISE) for the DOE via Oak Ridge
Associated Universities (ORAU) under DOE contract number DE-SC0014664;
G.W.H., A.H. and M.F. were supported by the Partnership for Multiscale Gyrokinetic Turbulence (MGK) and the High-Fidelity Boundary Plasma Simulation (HBPS) projects, part of the U.S. Department of Energy (DOE) Scientific Discovery Through Advanced Computing (SciDAC) program, via DOE contract DE-AC02-09CH11466 for the Princeton Plasma Physics Laboratory. Computations were performed on the Stellar cluster at Princeton University and the Cori cluster at NERSC. All opinions expressed in this paper are the authors'
and do not necessarily reflect the policies and views of DOE, ORAU, or ORISE.

\appendix
\section{Getting \gke~and reproducing results} \label{sec:getGkeyll}

Readers may reproduce our results and also use \gke~for their applications. The code and input files used here are available online. Full installation instructions for \gke~are provided on the \gke~website (\url{https://gkeyll.readthedocs.io}). The code can be installed on Unix-like operating systems (including Mac OS and Windows using the Windows Subsystem for Linux) either by installing the pre-built binaries using the conda package manager (\url{https://www.anaconda.com}) or building the code via sources. The input files used here are under version control and can be obtained from the repository at \url{https://github.com/ammarhakim/gkyl-paper-inp/tree/master/2022_PPCF_EMGK_shear}.

\bibliographystyle{iopart-num}

\bibliography{library}

\end{document}